\documentstyle [aps]{revtex}
\def\btt,1{{\tt$\backslash$,1}}
\begin{document}
%
\title{
Momentum dependence of the spin and charge excitations
in the two dimensional Hubbard model}
\author{ M. P. L\'opez-Sancho, F. Guinea and J. A. Verg\'es}
\address{
Instituto de Ciencia de Materiales de Madrid (Sede B). \\
Consejo Superior de Investigaciones Cient\'{\i}ficas. \\
Cantoblanco. E-28049 Madrid. Spain. }
\author{E. Louis}
\address{
Departamento de F\'{\i}sica Aplicada. \\
 Universidad de Alicante. \\
Apartado 99. E-03080. Alicante. Spain }
\date{\today}
\maketitle
\begin{abstract}
	The low energy spin and charge excitations in the 2D
Hubbard model near half filling are analyzed. The RPA spectra derived
from inhomoheneous mean field textures are analyzed.
Spin excitations show a commensurate peak at half filling,
incommensurate peaks near half filling, and a broad background typical
of a dilute Fermi liquid away from half filling.
Charge excitations, near half filling,  are localized near (0,0),
and they occupy a small portion of the Brillouin Zone, in
a way consistent with the existence of a small density of carriers,
and a small Fermi surface.
At higher hole densities, they fill the entire BZ, and can be understood
in terms of a conventional Fermi liquid picture.  The results are
consistent with the observed features of the high-T$_c$ superconductors.
\end{abstract}
	Despite its simplicity, the Hubbard model
in two dimensions is poorly understood\cite{sanse}, although it is widely
considered that it is a good description of the low energy
physics of high-T$_{\hbox{c}}$ superconductors\cite{phystod}.

	Among other possibilities, it has been conjectured that,
near half filling, it may show non trivial (marginal) behavior,
without coherent quasiparticles\cite{marginal}, or that it is a strongly
correlated d-wave superconductor\cite{dwave}. One of the main
stumbling blocks encountered in the study of this model is the
absence of a scheme general enough to describe such widely different
possibilities. Numerical calculations have been inconclusive
with respect to the low energy behavior in the lightly doped, strongly
coupled regime\cite{numer}.

	In the present work, we show that a global picture
can be obtained from mean field calculations (unrestricted
Hartree Fock) supplemented by the next leading order
correction, given by the Random Phase Approximation.
Note that such an approach does reproduce
correctly the main features of the Hubbard model in the regimes
where a consensus about its behavior has been reached:

	i) It describes a Luttinger liquid in 1D.

	ii) It gives an antiferromagnetic insulator at half filling
in a bipartite lattice.

	iii) It reproduces Fermi liquid behavior at low electron densities.

	iv) It satisfies Nagaoka's theorem for sufficiently low doping and
large values of $U/t$.

	It is commonly accepted that a mean field description suffices to
understand regimes ii), iii) and iv). In all cases, the ground state
wavefunction is well approximated by a Slater determinant (although, in order
to obtain correctly the low energy excitations of an antiferromagnet,
the RPA needs to be used).

	What is not so widely appreciated
is that, with some hindsight, the main features of a Luttinger
liquid can also be inferred from such a {\it mean field + RPA fluctations}
approach. In fact, the spin and the charge RPA polarizabilities
of a 1D gapless Fermi system differ by a sign in the denominator,
( $\chi_{RPA} \sim \chi_0 / ( 1 \pm U \chi_0 )$ ). As, in 1D, the only
excitations are associated with poles in $\chi$, this non trivial difference
in sign shifts the pole  in the spin-spin channel with respect
to that in the charge-charge channel, leading to spin-charge separation.
In addition, it can be shown that the injection of additional
electrons leads to alterations in the self consistent Hartree Fock
wavefunctions. These changes, in turn, renormalize the
quasiparticle weight, which vanishes in 1D, and near half filling
in 2D\cite{qp}. Thus, this approach corretly identifies the
two characteristc features of a Luttinger liquid:
the absence of a quasiparticle peak at low energies, and the
separation of spin and charge\cite{LL}.

	We now study the spin and charge excitations
of the 2D Hubbard model, within the RPA. To do so, we use as starting point,
the self consistent solutions at finite dopings extensively discussed
earlier\cite{hf}. The most charateristic feature of these solutions
is that, except for half filling or for a very small number of
particles, they correspond to inhomogeneous spin and charge
textures. These wavefunctions, which break the translational symmetry
of the Hubbard hamiltonian, have always a lower energy than the
homogeneous ones. The same spontaneous breaking of translational symmetry
occurs when applying mean field techniques to 1D systems: it is
the origin of the well known SDW or CDW instabilities. This instability
reflects the failure of standard perturbation theory around
the unperturbed wavefunction.

	In principle, translational symmetry
can be restored by hybridizing equivalent textures, shifted by
a given lattice vector. This procedure is equivalent to project
the $\vec{k} = 0$ component of the initial wavefunction.
It was shown in\cite{polaron} that, in this way, an effective mass
for the local textures (^^ ^^ spin polarons ") can be obtained,
along with other non trivial features later confirmed by other
numerical techniques, like the tendency of the defects to hop
within a given sublattice only. Analogously, rotational symmetry
can be retrieved by projecting the $S = 0$ component of the
inhomogeneous wavefunction. Note that, for the half filled case,
the translational and rotational symmetry breaking present
in the Hartree Fock wavefunction reproduces the accepted features
of the exact solution.

	In the present work, however, we have not tried to
restore translational symmetry. The RPA response functions, instead,
are analyzed in Fourier space. We obtain an approximation
to matrix elements of the type $\langle 0 | {\hat{A}}_{\vec{q}} | n \rangle$,
where $\hat{A} = \vec{S} , \hat{Q} $.
It was shown in\cite{polaron}
the hybridization effects do not change significantly the
main features of the wavefunctions, except for giving a
finite hopping amplitude to the local defects. Thus, we expect
the lack of translational invariance to be a higher order effect.
Our calculations also depend on the choice of spin texture,
for a given filling and value of $U/t$. As discussed in\cite{hf},
the Hartree Fock equations admit a variety of nearly degenerate,
inhomogeneous, self consistent solutions, and it is almost an
impossible task to identify the optimal one. It is likely that, in addition,
this richness of solutions is physically meaningful, being
associated to the intrinsic frustration induced by the
competition between conmensurability and kinetic energy
minimization. In the following,
we use ^^ ^^ spin polaron " solutions, which describe best
the low energy states for intermediate to large values
of $U/t$, and a wide range of dopings. These textures
correspond to solutions in which holes are localized in a
given site, in which the spin is flipped and reduced,
with the remaining structure retaining its antiferromagnetic
order. We have checked that
the main features reported below are insensitive with respect
to the type of texture being considered.

	The analysis performed here was described in\cite{RPA}.
We calculate:

\begin{equation}
	\chi ( \omega ) = \left [ {\cal I} \pm U {\chi}_0 ( \omega ) \right
        ]^{-1} {\chi}_0 ( \omega )
\label{chi}
\end{equation}

	where $\chi$ is written, in real space, as a $4 N \times 4 N$
matrix, for a system of size $N \times N$. The four components per site
correspond to the three spin degrees of freedom ( - sign in Eq.(\ref{chi}) )
and the charge degree of freedom ( + sign in Eq.(\ref{chi}) ).

	Once $\chi$ has been obtained in real space, we define:

\begin{equation}
	\chi ( \vec{k} , \omega ) = \sum_{i,j} e^{i \vec{k} ( \vec{r}_i
	- \vec{r}_j )} \chi_{i,j} ( \omega )
\end{equation}

	Our results allows us to identify four distinct regimes,
as function of doping and coupling:

	i) At half filling (one electron per site, n=1)
the excitations correspond to
those of a conmensurate antiferromagnetic insulator. The only
low energy excitations are the transverse spin modes, which
are dominated by the conmensurate peak at
$( \pi , \pi )$ (fig 1). Charge modes have no weight at low energies,
showing a gap of order $U$. This is an obvious case where the
scheme used here approximates well the expected features of the exact
solution. Note that, in this situation, the spin and charge
excitations are clearly separated.

	ii) Near half filling, the spin and charge excitations
show very different, almost opposite, dependence on $\vec{k}$.
The spin excitations move away from the $( \pi , \pi )$ position,
giving rise to four inconmensurate peaks at finite energies (figure 2a).
We find, in addition, the remnants of the conmensurate peak,
which dissappear at higher energies.
Note that, for the range of $U/t$ values considered, these extra peaks
cannot be ascribed to imperfect nesting. They are deformations of
the AF background, more similar to the effects found in the
lightly doped $t - J$ model. It is remarkable that these peaks
survive the disorder in the textures used as starting point.
The charge excitations (figure 2b), on the other
hand, have peaks near $\vec{k} = ( 0 , 0 )$. They arise
from the excitation of structureless states close to the Fermi
energy. They resemble the excitations in a low density
gas of electrons ( or holes ), also in agreement with the
expected results in the $t - J$ model. As in the previous case,
spin and charge excitations show almost non overlapping
features in the $( \vec{k} , \omega )$ space.
In addition, this regime corresponds
to the vanishing of the quasiparticle pole, also within
the unrestricted Hartree Fock approximation\cite{qp}.
These features are also present in the 1D Luttinger liquid state,
and are not expected in a conventional Fermi liquid,
described in terms of Landau's quasiparticles.
The region where this behavior can be clearly identified
is $n \ge 0.9$.

	iii) Our results indicate the existence of a smooth crossover
between the regime described above and a conventional
Fermi liquid al low dopings.
In this region, which covers most of
filling fractions, the spectral strength of the low
energy spin and charge excitations fall in
different regions of the BZ ( figures 3a and 3b ).
On the other hand, the eigenvectors of $\chi ( \omega )$
show a mixture of spin and chagre features.
In general,
this situation corresponds to the existence of a weak spin and
charge texture in the initial Hartree-Fock configuration.
In many cases, this texture is a well defined spin density wave.

	iv ) In a sufficiently dilute system, our results are
consistent with the existence of a normal Fermi liquid phase.
The spin and charge background is uniform, and the local
magnetization is zero everywhere. The RPA excitations show
a $6 \times 6$ structure (figures 4a and 4b), which is due to finite lattice
effects. This pattern is the same in the spin and charge channels,
in agreement with the absence of spin-charge separation.
We can establish the existence of this regime for
filling fractions $n \le 0.5$, for $U/t = 5$.

	It is interesting to compare our results with the growing
literature on the question of spin-charge separation in 2D.
Almost all schemes concentrate on one electron properties,
like the momentum distribution or the self energy.

	Near half filling numerical and analytical calculations support
the conclusion that a single hole in an antiferromagnetic background
shows no quasiparticle pole\cite{hole}.
Mean field results, along the approach
discussed here, confirm this conclusion\cite{qp}. In the latter case, the
analysis allows a rather straightforward description in terms of the
orthogonality catastrophe\cite{ortho}.
This is consistent with the work reported here,
as the low energy excitations do not exhibit an abrupt discontinuity
from the half filled case, where the spin-charge separation is obvious.

	Away from half filling existing approaches can be classified into
numerical calculations\cite{numersep1,numersep2},
trial wavefunctions\cite{trial}  and diagrammatic
analysis\cite{SCF}. Numerical evidence is difficult to interpret.
Exact results
for finite systems\cite{numersep2}
have been used in support of spin-charge separation,
based on the fact that the electronic momentum distribution resembles
that of spinless electrons.  Away from half filling, however, no evidence
for spin-charge separation is found\cite{numersep1}.
Alternatively, trial Luttinger-liquid-like wavefunctions\cite{trial}, also
support the existence of spin charge separation. Finally, self consistent
diagrammatic schemes lead to self energies which do not show the
expected Fermi liquid behavior\cite{SCF}. Thus, there exists a consensus that,
near half filling, the Hubbard model does not behave like a
conventional Fermi liquid.

	Our results corroborate the evidence discussed above.
Moreover, they present a qualitative picture of the low energy
spin and charge excitations, which clarifies the origin of the
separation between the spin and the charge degrees of freedom.
It should facilitate direct comparison with more phenomenological approaches
based on the existence of strong short range antiferrromagnetic
correlations\cite{phenom}, which are being used to explain d-wave
superconductivity in the high-T$_{\hbox{c}}$ superconductors.
The basic ingredient of these schemes is the existence of
local moments aligned antiferromagnetically. That is indeed
a feature present in our results.

	Finally, it is worth mentioning that our results agree
well with the measured magnetic properties of the high-T$_{\hbox{c}}$
cuprates\cite{exp}. It would be
interesting to check whether our predictions for
the charge excitations can also be observed.


\figure{Figure 1. Transverse spin excitations of the Hubbard model
in a square lattice ( $12 \times 12$ sites )
 for $U/t = 10$, $n = 1$ (half filling) and
$\omega = 0.1 t$. $k_x$ and $k_y$ are plotted in units of
$\pi / a$, where $a$ is the lattice constant.}

\figure{Figure 2a. Charge excitations, calculated for $U/t$ = 10,
$n$ = 0.9 and $\omega = 0.1$. Other parameters as in fig. 1.
Figure 2b. Spin excitations. }

\figure{Figure 3a. Charge excitations, calculated for $U/t$ = 10,
$n$ = 0.5 and $\omega = 0.1$. Other parameters as in fig. 1.
Figure 3b. Spin excitations. }

\figure{Figure 4a. Charge excitations, calculated for $U/t$ = 5,
$n$ = 0.3 and $\omega = 0.1$. Other parameters as in fig. 1.
Figure 4b. Spin excitations. }


\begin{references}

\bibitem{sanse}
See, for instance, the contributions in
{\bf The Physics and Mathematical Physics of the
Hubbard Model}, D. Baereswyl, J. Carmelo, D. K. Campbell, F. Guinea
and E. Louis eds., Plenum Press (New York), 1994.

\bibitem{phystod}
P. W. Anderson and R. Schrieffer, Physics Today
{\bf 44} (June), 54 (1991).

\bibitem{marginal}
C. M. Varma, P. B. Littlewood, S. Schmitt-Rink, E. Abrahams
and A. E. Ruckenstein, Phys. Rev. Lett. {\bf 63}, 1996 (1989).

\bibitem{dwave}
C.-H. Pao and N. E. Bickers, Phys. Rev. Lett. {\bf 72}, 1870 (1994).
P. Monthoux and D. J. Scalapino, {\it ibid} {\bf 72}, 1874 (1994).
K. Maki and H. Won, {\it ibid} {\bf 72}, 1758 (1994).

\bibitem{numer}
M. Fabrizio, A. Parola and E. Tosatti
Phys.\ Rev.\ B {\bf46}, 3159 (1992).
E. Dagotto, A. Moreo, F. Ortolani, D. Poilblanc and J. Riera,
Phys.\ Rev.\ B {\bf 45}, 10741 (1992).

\bibitem{LL}
P. D. M. Haldane, J. Phys. C {\bf 14}, 201 (1981).

\bibitem{qp}
G. Gal\'an, F. Guinea, J.A. Verg\'es, G. Chiappe and E. Louis,
Phys.Rev.B {\bf 46}, 3163 (1992); E. Louis, G. Chiappe, J. Gal\'an,
F. Guinea and J.A. Verg\'es, Phys. Rev. B {\bf 48}, 426 (1993).

\bibitem{hf}
A. R. Bishop, F. Guinea, P. S. Lomdahl, E. Louis and J. A. Verg\'es,
Europhys. Lett. {\bf 14}, 157 (1991).
J. A. Verg\'es, E. Louis, P. S. Lomdahl, F. Guinea and A. R. Bishop,
Phys. Rev. B {\bf 43}, 6099 (1991).

\bibitem{polaron}
E. Louis, G. Chiappe, F. Guinea, J. A. Verg\'es and E. V. Anda, Phys. Rev
B {\bf 48}, 9581 (1993).

\bibitem{RPA}
F. Guinea, E. Louis and J. A. Verg\'es,
Phys. Rev. B {\bf 45}, 4752 (1992).
F. Guinea, E. Louis and J. A. Verg\'es,
Europhys. Lett. {\bf 17}, 455 (1992).

\bibitem{hole}
Q. F. Zhong, S. Sorella and A. Parola, Phys. Rev. B {\bf 49}, 6408 (1994).
E. M\"uller-Hartmann
and C. I. Ventura, Phys. Rev. B {\bf 50}, 9235 (1994).

\bibitem{ortho}
P. W. Anderson,
Phys. Rev. {\bf 164}, 352 (1967).

\bibitem{numersep1}
E. A. Jagla, K. Hallberg and C. A. Balseiro,
Phys. Rev. B {\bf 47}, 5849 (1993).

\bibitem{numersep2}
M. Arjunwadkar, P. V. Panat and D. G. Kanhere, Phys. Rev. B
{\bf 48}, 10563 (1993).
Y. C. Chen, A. Moreo, P. Ortolani, E. Dagotto and T. K. Lee,
Phys. Rev. B {\bf 50}, 655 (1994).
W. O. Putikka, R. L. Glenister, R. R. P. Singh and H. Tsunetsugu,
Phys. Rev. Lett. {\bf 73}, 170 (1994).
Y. C. Chen and T. K. Lee,  Zeits. f. Phys. B {\bf **}, **** (1994).

\bibitem{trial}
R. Valenti and C. Gros, Phys. Rev. Lett. {\bf 68}, 2402 (1992) and
Phys. Rev. B {\bf 50}, 11313 (1994).

\bibitem{SCF}
N. E. Bickers and S. R. White, Phys. Rev. B {\bf 43}, 8044 (1991).
See also\cite{dwave}.
Y. M. Yilk, L. Chen and A.-M. S. Tremblay, Phys. Rev B {\bf 49}, 13267 (1994).

\bibitem{phenom}
P. Monthoux and D. Pines, Phys. Rev. B {\bf 49}, 4261 (1994).


\bibitem{exp}
J. Rossat-Mignod {\it et al}, Physica {\bf B169}, 58 (1991).
T. E. Mason {\it et al}, Phys. Rev. Lett. {\bf 68}, 1414 (1992).

\end{references}
\end{document}